\documentclass[pra,aps,showpacs,amssymb,amsmath]{revtex4}
\usepackage{latexsym}
\def\imag{i}
\begin{document}

\title{Polarization tensors and the photon field}
\author{Brian Seed}
\affiliation{Center for Computational and Integrative Biology, Wellman 911, Massachusetts General Hospital, Boston MA 02114 USA}

\begin{abstract}
A direct calculation of the elements of the photon polarization vector for arbitrary momentum in the helicity basis shows that it is not a vector but a complex bivector. The bivector real and imaginary parts can be directly equated with electromagnetic field amplitudes and the associated field equations are the Maxwell equations in time-imaginary space. The bivector field exhibits a phase freedom (Berry, or geometric phase) dependent on the rotation history of the field or observer. Phase freedom is not intrinsically present in the longitudinal excitations of the field and a general argument connects quantization of angular momentum with the observation of phase changes associated with frame rotation. Current and translation operators can be defined for bivector fields that are free of defects associated with a quantized vector potential.

\bigskip

\end{abstract}
\maketitle

\section{Introduction}
Wigner first pointed out that quantum states should be representations of the inhomogeneous Lorentz group \cite{Wigner39}. However application of this simple intuitive principle has proven surprisingly challenging, particularly in the context of the massless spin one field \cite{Weinberg95}. One of the most widely encountered representations of the photon field is of a quantized Gibbs (column) vector potential 
\begin{equation}\label{standard_field}
A^{\,\mu}(x)= (2\pi)^{-3/2} \sum_\sigma \int^\infty_{-\infty} \frac{d^3 p}{\sqrt{2p^0}}[e^{\mu}(\mathbf{p},\sigma) a(\mathbf{p},\sigma)e^{\imag p x} + e^{\mu *}(\mathbf{p},\sigma) a^\dagger(\mathbf{p},\sigma)e^{-\imag p x}]
\end{equation}
in which $e^{\mu}(\mathbf{p},\sigma)$, the photon polarization vector, satisfies $p_{\mu}e^{\mu}(\mathbf{p},\sigma)=0, e^0(\mathbf{p},\sigma)=0$. For a pure motion in the 1-direction 
\begin{equation}\label{pure1directionbivector}
e(\mathbf{p},\pm \sigma) = (0,0,\pm \imag,1)/\sqrt{2}
\end{equation}
Because the polarization vectors are transverse, they impart to the field the influence of holonomy, a quantity associated with the change in orientation of a tangent vector directed by parallel transport around a non-Euclidean surface. A substantial literature has developed around the discovery by Berry \cite{Berry84} that the transport of a quantum system through an adiabatic circuit induces a change in phase that is dependent on the area of the surface enclosed and the angular momentum of the system. The identification of this effect with holonomy \cite{Simon83}, its realization experimentally through experiments with light \cite{Chiao86,Tomita86}, the elucidation of its origins in the Poincar\'{e} group \cite{Bialynicki-Birula87,Jordan87,Lindner03}, and the ensuing debate over whether the experimentally observed effects have classical or quantum origins \cite{Haldane86,Haldane87,Segert87,Berry87Nature} have recently been reviewed comprehensively \cite{Ben-Aryeh04}. Berry's original treatment has been extended to non adiabatic circuits \cite{Aharanov87} and open trajectories \cite{Samuel88,Jordan88}; and the contributions of holonomy were recognized \cite{Ramaseshan86} to be implicit in earlier studies of polarization phase by Pancharatnam \cite{Pancharatnam56,Berry87JModOpt} and in the proposal by Aharanov and Bohm \cite{Aharanov59} to measure phase changes in non-simply connected topologies. 

In this work we carry out a direct calculation to demonstrate that the photon polarization `vector' is not a vector but a complex bivector, and that the bivector can be equated with the electromagnetic field, not the vector potential. The bivector field exhibits a phase freedom associated with rotation history that cannot be separated from the observer frame. If the widely used Gibbs vector formalism is extended to spaces in which the time dimension has imaginary weight, a covariant Hamiltonian can be created that yields well-behaved eigensolutions with positive and negative energy eigenvalues and that yields the Maxwell equations upon application to the fields. Current and translation operators can be defined that are free of various defects associated with the quantized vector potential. Experimental tests can be devised that distinguish between a vector potential and a bivector, using a simple idealization of a single mode state to explore contrasting predictions of the dependence of the energy of a photon on its modulation envelope. 

\section{A covariant operator algebra for Gibbs vectors}
In the following the metric signature is $(-+++)$ and `a pure motion in the $n$-direction' signifies that the velocities along the spatial axes $j\neq n$ are null. Our first task is to develop a covariant Gibbs vector algebra that allows a Hamiltonian equation for the electromagnetic field. To achieve this we introduce operators $\chi^{\,\mu}$ that can be considered analogs, in a vector representation, of the Weyl operators $\sigma^{\,\mu}$ in the spinor representation. The components $\chi^{\,\mu}$ (formally, ${\chi^{\,\mu\nu}}_\rho$) are rank two tensors that can be written
\begin{equation}\label{chi_matrices}\{ \left(\begin{matrix} 1 & 0 & 0 & 0 \cr 0 & 1 & 0 & 0 \cr 0 & 0 & 1 & 0 \cr 0 & 0 & 0 & 1 \cr  \end{matrix}\right) ,
  \left(\begin{matrix} 0 & \imag  & 0 & 0 \cr -\imag  & 0 & 0 & 0 \cr 0 & 0 & 0 & -\imag  \cr 0 & 0 & \imag
     & 0 \cr  \end{matrix} \right) ,\left(\begin{matrix} 0 & 0 & \imag  & 0 \cr 0 & 0 & 0 & \imag  \cr -\imag
      & 0 & 0 & 0 \cr 0 & -\imag  & 0 & 0 \cr  \end{matrix}\right)  ,
  \left(\begin{matrix} 0 & 0 & 0 & \imag  \cr 0 & 0 & -\imag  & 0 \cr 0 & \imag  & 0 & 0 \cr -\imag
      & 0 & 0 & 0 \cr  \end{matrix}\right)  \}
\end{equation}
The ${\chi^{\,\mu\nu}}_\rho$ and the related ${{\chi^{\,\mu}}_\nu}^\rho$ constitute two subalgebras of $Mat(4,\mathbb{C})$ that are isomorphic to the left and right subalgebras of the real matrix representation of quaternion multiplication $Mat(4,\mathbb{R})\simeq \mathbb{H}\otimes\mathbb{H}$ \cite{Lounesto01}. The three spatial operators $\chi^j$ have eigenvalues of $\pm 1$ and eigensolutions that are irreducible representations of the Lorentz group in a space that assigns imaginary weight to the (scalar) time direction. In this space the boost transformations are reflections, so the fundamental group operation is a reflection. Reflections are in any case more primitive than rotations since by a theorem of Hamilton any rotation can be decomposed as a pair of reflections, e.g. \cite{Cartan66}. The normalized eigensolutions of $\chi_1$ are:
\begin{equation} \label{eigenvectors1}
\{ \frac{1}{\sqrt{2}}\left( \begin{matrix} 0\cr 0\cr -\imag\cr 1\end{matrix}\right) ,\,\frac{1}{\sqrt{2}}\left( \begin{matrix} 0\cr 0\cr \imag\cr 1\end{matrix}\right) 
,\,\frac{1}{\sqrt{2}}\left( \begin{matrix} \imag\cr 1\cr 0\cr 0\end{matrix}\right) ,\,\frac{1}{\sqrt{2}}\left( \begin{matrix} -\imag\cr 1\cr 0\cr 0\end{matrix}\right)  \} 
\end{equation}
with respective eigenvalues $\{ 1, -1, 1, -1 \}$. The first pair are eigenbivectors or simply bivectors and correspond to the two orthogonal states of the polarization `vector' (\ref{pure1directionbivector}); the second pair are time-imaginary vectors. Like the $\sigma^{\,j}$, the $\chi_{\,j}\, , j \in {1,2,3} $ form the basis of a Clifford or geometric algebra $\mathcal{C}l_3$ on $\mathbb{R}^3$ defined in the usual way by
\begin{equation} \label{clifford_algebra}
{\chi_i}\,{\chi_j} + {\chi_j}\,{\chi_i} = 2 \delta_{\,ij} \mathbf{1}_4
\end{equation}
where $\mathbf{1}_4$ is the $4\times 4$ identity matrix. By (\ref{clifford_algebra}) appropriately normed Clifford numbers composed from the $\chi^{i}$ are elements of the (odd + even) group $\mathbf{Pin}(3)$ with $\mathbf{Pin}(3)/\{\pm1\} \simeq SO(3)$, since $\mathbf{Pin}(n)$ forms a double cover of $SO(n)$, \cite{Lounesto01,Porteous95}. The $\chi^{i}$ can be divided into spatial and temporal parts $\chi^{i} = K^{\,i} + S^{\,i} = K_{\,i} + S_{\,i}$, where 
\begin{equation}
{{S_0}^j}_k=\delta_{jk}\quad\quad {{S_i}^0}_\mu = {{S_i}^{\,\mu}}_0 = 0 \quad\quad {{S_i}^j}_k = -\imag \epsilon_{i\,j\,k}
\end{equation}
\begin{equation}
{{K_0}^0}_0=1\quad\quad {{K_i}^0}_j = -{{K_i}^{\,j}}_0 = \imag\, \delta_{ij}
\quad\quad 
{{K_\mu}^j}_k = 0
\end{equation}
The $S_i$ are the generators of the spin one rotation operator about axis $i$, often written ${{\mathfrak{J}_i}^j}_k$, and have, with the $K_i$, the following properties:
\begin{equation} 
{{S_i}}^2 +{{K_i}}^2 := {{S_i}^j}_k\,{{S_i}^k}_l + {{K_i}^j}_k\,{{K_i}^k}_l  = {\delta^j}_l
\end{equation}
\begin{equation}
e^{\kappa \,{S_i}} = {{K_i}}^2 + {{S_i}}^2\, \cosh \,\kappa + {S_i}\,\sinh \,\kappa 
\end{equation}
The $S_i$ are also the generators of bivector boosts in direction $i$. The $K_i$ are the generators of vector boosts in direction $i$ and satisfy
\begin{equation}
e^{\kappa \,{K_i}} = {{S_i}}^2 + {{K_i}}^2\,\cosh \,\kappa  + {K_i}\,\sinh \,\kappa
\end{equation}
The first two and last two elements of (\ref{eigenvectors1}) are eigen(bi)vectors of $S_1$ and $K_1$ respectively.

The $S_i$ and $K_i$ obey the commutation relations
\begin{eqnarray*} 
&\left[ {S_i}\,,{S_j}\right] & = \imag \,\epsilon _{ij\,k}\,S_k \\
&\left[ {K_i}\,,{K_j}\right] & = \imag \,\epsilon _{ij\,k}\,S_k \\
&\left[ {S_i}\,,{K_j}\right] & = \imag \,\epsilon _{ij\,k}\,K_k
\end{eqnarray*}
and the above relations differ from the usual formulation only in the sign of the righthand side of the middle equation. However as a result the operators $\chi_i$ conform to the Lie algebra of the quaternions, 
\begin{equation}\label{structure_equations}
\left[ {\chi_i}\,,{\chi_j}\right]  = 2\,\imag \,\epsilon _{ij\,k}\,\chi_k
\end{equation}

The Hamiltonian operator $\chi^j p_j$ has doubly degenerate eigenvalues $\pm p_0$ and hence admits a covariant equation of the form
\begin{equation}\label{covariant_equation}
\chi^{\,\mu} p_\mu \psi = 0
\end{equation} 
as well as a contravariant (opposite helicity) equation: 
\begin{equation}\label{contravariant_equation}
\chi^{\,\mu} p^{\,\mu} \psi = 0
\end{equation} 

In the case of a pure 1-direction motion the first eigenbivector of (\ref{eigenvectors1}), $\psi_+=( 0, 0, -\imag , 1)/\sqrt{2}$ is a solution of (\ref{covariant_equation}) and the second, $\psi_- =( 0, 0, \imag  , 1)/\sqrt{2}$ is a solution of (\ref{contravariant_equation}), and these correspond to the positive energy solutions for a positive and negative helicity particle, respectively. To the $\psi_\pm$ can be associated complex conjugates $\overline{\psi}_\pm$; their modulus, or length, $\overline{\psi}_\pm\cdot \psi_\pm$ is unity. Like their counterparts in the massless spin 1/2 field, $\psi_+$ and $\psi_-$ undergo boost transformations by different laws, but behave the same way under rotations. The $\psi_\pm$ can be rotated to an arbitrary direction of propagation by ${\psi_\pm}' = e^{-\imag \lambda_2 \, S_2}e^{-\imag \lambda_3 \,S_3}\psi_\pm$ where $\lambda_2 = \arcsin(-p_3/\sqrt{p^2_1+p^2_3})$ and $\lambda_3 = \arcsin(p_2/\sqrt{p^2_1+p^2_2+p^2_3})$, the order of multiplication of the rotation operators being given by the leftmost first, rightmost last prescription (or in a spinor context, outermost first, innermost last), see e.g. \cite{Cartan66} sec. 39. This gives
\begin{eqnarray}
\label{z2}
z_2(p) := e^{-\imag \lambda_2 \, S_2}e^{-\imag \lambda_3 \,S_3}\psi_+ =
\frac{( 0,\imag \,p_1\,p_2 - p_0\,p_3, -\imag \,\left( {p_1}^2 + {p_3}^2 \right),p_0\,p_1 + \imag \,p_2\,p_3)}{\sqrt{2}\,p_0\,\sqrt{{p_1}^2 + {p_3}^2}}
\end{eqnarray}

Similar actions on pure 2- and 3-direction bivectors, $\phi_+ = ( 0, 1, 0, -\imag )/\sqrt{2}$ and $\xi_+ = ( 0, -\imag, 1, 0)/\sqrt{2}$ respectively, lead to
\begin{equation} 
\label{z3}
z_3(p) :=
e^{-\imag \mu_3 S_3}e^{-\imag \mu_1 S_1}\phi_+ =
\frac{(0,{p_0}\,{p_2} + \imag \,{p_1}\,{p_3},
  -{p_0}\,{p_1}  + \imag \,{p_2}\,{p_3} ,
 -\imag \,\left( {p_1}^2 + {p_2}^2 \right) 
    )}{\sqrt{2}\,p_0\,\sqrt{\left( {p_1}^2 + {p_2}^2 \right)}}  
\end{equation} 
where $\mu_3 = \arcsin(-p_1/\sqrt{p^2_1+p^2_2}$ and $\mu_1 = \arcsin(p_3/\sqrt{p^2_1+p^2_2+p^2_3})$ and 
\begin{equation} 
\label{z1}
z_1(p) := 
e^{-\imag \nu_1 S_1}e^{-\imag \nu_2 S_2}\xi_+
=\frac{( 0,-\imag \,\left({p_2}^2 + {p_3}^2 \right),\imag \,p_1\,p_2 + {p_0}\,{p_3},- {p_0}\,{p_2}  + \imag \,{p_1}\,{p_3})}{{\sqrt{2}}\,p_0\,\sqrt{\left({p_2}^2 + {p_3}^2 \right) }} 
\end{equation}
where $\nu_1 = \arcsin(-p_2/\sqrt{p^2_2+p^2_3})$ and $\nu_2 = \arcsin(p_1/\sqrt{p^2_1+p^2_2+p^2_3})$. A structure closely related to $z_3$ can be found as a column element of a unitary matrix governing transformation of spin one states \cite{Bialynicki-Birula87}. 

The initial values of $\psi_+$, $\zeta_+$ and $\xi_+$ above have been chosen for consistency under rotation of the coordinate axes between the positive helicity bivector solutions and the positive momentum vector solutions
\begin{equation} \label{eigenvectors}
(\imag,1,0,0),(\imag,0,1,0) ,(\imag,0,0,1)
\end{equation}
The forward cycle $(1 \rightarrow 2 \rightarrow 3 \rightarrow 1)$ of transformations of both the bivectors and the vectors are produced by successive action of the axis rotation operator 
$e^{\frac{-2\,\pi\,i \,\left( {S_1} + {S_2} + {S_3} \right) }{3\,\sqrt{3}}}$.
The reverse cycle $(1 \leftarrow 2 \leftarrow 3 \leftarrow 1)$ is generated by the operator formed from the inverse (= the square) of the positive cycle operator.

Despite their apparent dissimilarities, each of the representations (\ref{z2} - \ref{z1}) is equivalent; each has unit length, gives rise to energy-momentum currents of the same magnitude (discussed below) and transforms under boosts and rotations in the same way. However each $z_j$ is poorly defined for a pure motion in direction $j$. The normalized sum of bivectors
\begin{equation} \label{z_S}
z_S(p) := \frac{{\sqrt{{{p_2}}^2 + {{p_3}}^2}}\,{z_1(p)} + {\sqrt{{{p_1}}^2 + {{p_3}}^2}}\,{z_2(p)} + {\sqrt{{{p_1}}^2 + {{p_2}}^2}}\,{z_3(p)}}
  {{\sqrt{{\left( {p_1} - {p_2} \right) }^2 + {\left( {p_2} - {p_3} \right) }^2 + {\left( -{p_1} + {p_3} \right) }^2}}} 
\end{equation}
with components
\begin{eqnarray} \label{z_S_components}
z_S(p)_0 = & 0 \nonumber\\
z_S(p)_j = & \frac{\sum_{k = 1}^{3}\,\left( {-p_0}\epsilon _{jkm}\,{p_m} + \imag \,( {p_j}\,{p_k}   - {{p_k}}^2 )\right)}{{\sqrt{2}}\,{p_0}\,{\sqrt{{\left( {p_1} - {p_2} \right) }^2 + {\left( {p_1} - {p_3} \right) }^2 + 
        {\left( {p_2} - {p_3} \right) }^2}}} 
\end{eqnarray}
(with implicit summation over $m$) is free of this defect. The subscript $S$ reflects the space-like character of the polarization vector. 

Upon rotation to arbitrary momentum the negative eigenvalue solutions in (\ref{eigenvectors1}) yield the complex conjugates of (\ref{z2} - \ref{z1}) and (\ref{z_S}), corresponding to the solutions for negative helicity and denoted by overlining. Because
\begin{equation}
\overline{z}_S \,z_S  = z_S \, \overline{z}_S = 1 \quad\quad
z_S \,z_S = \overline{z}_S \,\overline{z}_S = 0
\end{equation}
the $z_S$ and $\overline{z}_S $ recapitulate the familiar orthogonality relations of the polarization vector for distinct helicities $\sigma, \sigma'$:
\begin{equation}
{e^*}_\mu(\mathbf{p},\sigma')e^{\,\mu}(\mathbf{p},\sigma) = \delta_{\sigma,\sigma'}
\end{equation}
In a similar manner we find that 
\begin{eqnarray}\label{projection_matrix}
\sum_\sigma e_\mu(\mathbf{p},\sigma){e^*}_\nu(\mathbf{p},\sigma) = & z_S \otimes \overline{z}_S +\overline{z}_S \otimes z_S \nonumber \\
& {\delta_{\mu\,\nu}} -{p_\mu\,p_\nu}\,/{p_0}^2  & 1\leq \mu,\nu \leq 3 \nonumber \\
 = & 0 & \textrm{otherwise} 
\end{eqnarray}
which is the standard relation for the projection matrix of the polarization vector. 

\section{Identity with Electromagnetic Field}
The $z_j$ are solutions of (\ref{covariant_equation}), but describe a field that is not a simple vector potential. Using the correspondence between the symmetric field stress-energy tensor $\Theta^{\,\alpha \beta} := 
   -g^{\,\alpha \mu }\,{F_{\mu \lambda }}\,F^{\,\lambda \beta } - 
    \frac{1}{4}\,g^{\,\alpha \beta }\,{F_{\mu \lambda }}\,F^{\,\mu \lambda }$ and the stress-energy tensor $T^{\alpha \beta}$ with components ${p_\alpha p_\beta}/{p_0}$ to translate between momentum and field representations, we find
\begin{equation}\label{maxwell_field} z_j(p) = 
\frac{-\imag}{\sqrt{2 p_0}}\frac{-\imag \,{b_j} + {e_j}}{\sqrt{{b_j}^2 + {e_j}^2}}\left( 0, \imag \, {b_1} + {e_1}, \imag \, {b_2} + {e_2}, \imag \, {b_3} + {e_3}\right) 
\end{equation}
and 
\begin{equation}\label{normed_maxwell_field} z_S(p) = \frac{-\imag}{\sqrt{2 p_0}} \frac{\sum_j -\imag \,{b_j} + {e_j}}{\sqrt{(\sum_j{b_j})^2 + (\sum_j{e_j})^2}} \left( 0, \imag \, {b_1} + {e_1}, \imag \, {b_2} + {e_2}, \imag \, {b_3} + {e_3}\right) \end{equation}
The differences in representation reflect, as in the momentum formulation, alternative normalizations differing in phase. The normalization factor of $1/\sqrt{2 p_0}$ emerges naturally as a consequence of the unit length of the bivector. 

With (\ref{maxwell_field}) and (\ref{normed_maxwell_field}) we have found that explicit application of the program to represent particles as irreducible representations of the inhomogeneous Lorentz group in the Gibbs vector representation leads to the conclusion that the elementary excitation of the massless spin one field is not a vector but a complex bivector. In the normalized form of (\ref{z_S}), the real and imaginary parts can be equated with the electric and magnetic amplitudes of the photon field respectively. That is, (\ref{z_S}) can be decomposed as
\begin{equation}
z_S(p) := ze(p)+\imag\, zb(p)
\end{equation}
where, for example, 
\begin{eqnarray}\label{ze(p)_zb(p)}
ze(p)_1 = &  p_0 (p_2-p_3)/N \nonumber\\
zb(p)_1 = & (-{p_0}^2 + p_1 \sum_{j=1}^3 p_j )/N \nonumber\\
N = & {\sqrt{2}}\,{p_0}\,{\sqrt{{\left( {p_1} - {p_2} \right) }^2 + {\left( {p_1} - {p_3} \right) }^2 + 
        {\left( {p_2} - {p_3} \right) }^2}}
\end{eqnarray}
$ze(p)$ reverses upon spatial reflection ($p_i \rightarrow -p_i$) but not time reflection ($p_0 \rightarrow -p_0$), whereas $zb(p)$ shows the opposite behavior, consistent with the known symmetries of the electric and magnetic fields. The $ze(p)$ and $zb(p)$ constitute an orthogonal basis from which the projection matrix can also be derived.
\begin{equation}
 z_S \otimes \overline{z}_S +\overline{z}_S \otimes z_S = 2\,( ze(p) \otimes ze(p) + zb(p) \otimes zb(p) )
\end{equation}

With the results above we can define a positive helicity spin one field
\begin{equation}\label{psi}
\psi_+ = \frac{e^{\imag\,\theta}}{\sqrt{2p_0}}( 0,\imag \,b_1 + e_1,\imag \,b_2 + e_2,\imag \,b_3 + e_3) 
\end{equation}
in which the factor $e^{\imag\, \theta}$ assimilates the phase freedom of the field. 
The field equations
\begin{equation} \chi^{\,\mu} \partial_\mu  \psi_+ = (-\mathbf{\nabla \cdot b} + \imag \mathbf{\nabla \cdot e}, \partial_0 \mathbf{e} - \mathbf{\nabla \times b} + \imag (\partial_0 \mathbf{b} + \mathbf{\nabla \times e} ))\frac{e^{\imag \,\theta}}{\sqrt{2p_0}} = (\imag \rho, -\mathbf{j})\frac{e^{\imag\, \theta}}{\sqrt{2p_0}}
\end{equation}
are related to the classical Maxwell equations in a time-imaginary space. The negative helicity field, which is the complex conjugate, follows the equation
\begin{equation} \label{maxwell_negative}
\chi^{\,\mu} \partial^{\,\mu}  \psi_- = (\imag \rho, \mathbf{j})\frac{e^{-\imag\, \theta}}{\sqrt{2p_0}}
\end{equation}
so the conventional representation, (\ref{standard_field}), is not a solution of a single field equation and the field must be represented instead as a direct sum $\psi_+ \oplus \psi_-$. Application of the positive helicity equation to the negative helicity field or vice versa results in a mispairing of the sign of the time derivative with that of the cross product, i.e. yields equations of the form
\begin{equation}
\partial \mathbf{e} + \nabla \times \mathbf{b} =0\quad \mathrm{and} \quad \partial \mathbf{b} - \nabla \times \mathbf{e} = 0
\end{equation} 
The sourceless field equation associated with the direct sum can be written
\begin{equation} \label{direct_sum}
\partial_\mu\,\left(\begin{matrix} \chi^{\,\mu} & \mathbf{0}_4 \cr \mathbf{0}_4 & -\chi_{\,\mu}  \cr  \end{matrix}\right) \left(\begin{matrix} \psi_+\cr -\psi_-\cr \end{matrix}\right) := \mathcal{G}_\partial \,\left(\begin{matrix} \psi_+\cr -\psi_-\cr \end{matrix}\right)= 0
\end{equation}
$\mathcal{G}_\partial$ being the operator inner product, an $8\times 8$ matrix.

\section{Transformation properties of the fields}
The boost transformations of $z_S$ and $\overline{z}_S$ differ
\begin{equation}\label{z_S_boost}
z_S(p^\prime) = e^{- \kappa \, S_i} z_S(p)= z_S(p) e^{\, \kappa \, S_i}  \quad\quad \overline{z}_S(p') = e^{\,\kappa \, S_i} \overline{z}_S(p) =  \overline{z}_S(p) e^{-\kappa \, S_i}
\end{equation}
whereas the rotations are the same
\begin{equation}\label{z_S_rotation}
 z_S(p^\prime) = e^{\,\imag \kappa \, S_j} z_S(p)=  z_S(p) e^{-\imag \kappa \, S_j}\quad\quad  \overline{z}_S(p^\prime) = e^{\,\imag \kappa \, S_j} \overline{z}_S(p) = \overline{z}_S(p) e^{-\imag \kappa \, S_j} 
\end{equation}
Applied to the representation (\ref{direct_sum}) this implies that the generators of boosts and rotations have the forms 
\begin{equation}\label{generators}
\left(\begin{matrix} S_{i} & \mathbf{0}_4 \cr \mathbf{0}_4 & -S_{i}  \cr  \end{matrix}\right) \quad\mathrm{and}\quad \left(\begin{matrix} S_{i} & \mathbf{0}_4 \cr \mathbf{0}_4 & S_{i}  \cr  \end{matrix}\right)
\end{equation}
respectively. Although the Gibbs vector formalism is the most frequently encountered depiction of the electromagnetic field, the most compact way to represent the field operators is not by Gibbs bivectors but by spinor bivectors (or, more directly, by elements of a Clifford algebra \cite{Hestenes99,Baylis99,Lounesto01}), which will be presented elsewhere. The time and space reflection operators defined by
\begin{equation}
R_T :=\left(\begin{matrix} \mathbf{0}_4 & -\mathbf{1}_4 \cr \mathbf{1}_4 & \mathbf{0}_4  \cr  \end{matrix}\right) \quad\mathrm{and}\quad R_S := \left(\begin{matrix} \mathbf{0}_4 & \mathbf{1}_4 \cr \mathbf{1}_4 & \mathbf{0}_4  \cr  \end{matrix}\right) 
\end{equation}
have action on the operator $\mathcal{G}_\partial$ of (\ref{direct_sum}) given by the inner products $R_T \mathcal{G}_\partial  R_T$ and $R_S \mathcal{G}_\partial  R_S$ respectively, consistent with their character as reflections, not rotations. 

\section{Phase freedom and topology}
Because the lengths of the bivectors (\ref{z2} - \ref{z1}) are equal, the differences between them constitute variations in normalization that affect the argument but not modulus of a complex quantity, a form of phase freedom that can be identified with the well-studied Berry phase \cite{Berry84,Simon83,Berry87Nature}, recently reviewed in \cite{Ben-Aryeh04}. The phase has a history dependence which is easily displayed without reference to circuit as follows.

If the order of rotations is reversed for each of the (\ref{z2} - \ref{z1}), a different collection of bivectors results.  For example, from the pure 1-direction bivector $\psi_\pm$ we can carry out the rotation ${\psi_\pm}' = e^{-\imag \lambda_3 \, S_3}e^{-\imag \lambda_2 \,S_2}\psi_\pm$ where $\lambda_2 = \arcsin(-p_3/\sqrt{p^2_1+p^2_2+p^2_3})$ and $\lambda_3 = \arcsin(p_2/\sqrt{p^2_1+p^2_2})$. In this case the resulting ${\psi_\pm}'$ is poorly defined for motion in the 3-direction, not the 2-direction as in (\ref{z2}). The collection of bivectors under reversal of the rotation order then are:
\begin{equation}
{\psi_+} \rightarrow y_3, {\zeta_+} \rightarrow y_1, {\xi_+} \rightarrow y_2 
\end{equation}
with the translation to field components
\begin{equation}\label{alternate_maxwell_field}
y_j = \frac{1}{\sqrt{2 p_0}}\frac{-\imag \,{b_j} + {e_j}}{\sqrt{{b_j}^2 + {e_j}^2}}\left( 0, \imag \, {b_1} + {e_1}, \imag \, {b_2} + {e_2}, \imag \, {b_3} + {e_3}\right)
\end{equation}
differing from (\ref{maxwell_field}) by a factor of $\imag$. Thus we can easily calculate changes in phase without completing a circuit. Alternatively, any open path can be turned into at least one closed circuit by joining the endpoints by a geodesic, along which the angle of rotation by parallel transport is zero \cite{Jordan88,Samuel88}

By rotating from a pure motion in the 1-direction to a motion in an arbitrary direction via the sequence leading to (\ref{maxwell_field}), and then back to a pure motion in the 1-direction by the reverse of the sequence leading to (\ref{alternate_maxwell_field}), a rectilinear path is inscribed on a sphere, and the regenerated bivector at the close of the loop has the form
\begin{equation}\label{rotation_phase}
e^{\pm\imag \arctan (\frac{p_2 p_3}{p_0 p_1})} \psi_\pm
\end{equation}
The simplest interpretation of (\ref{rotation_phase}) is that the positive helicity bivector has been rotated by $\arctan (\frac{p_2 p_3}{p_0 p_1})$ and the negative helicity bivector has been rotated by $-\arctan (\frac{p_2 p_3}{p_0 p_1})$. However if a positive helicity state is rotated about the axis of propagation by an amount $\gamma$ and a negative helicity state is rotated about the same axis by $-\gamma$, the effect of rotation is equivalent to a translation in time of the phase of both states by the same amount. In particular, for any equally weighted mixture of such states the holonomy angle should be zero. This conclusion would be consistent with the inferred dependence of phase on angular momentum \cite{Berry84,Bialynicki-Birula87,Jordan87}, but leads to the prediction that a linearly polarized beam, which has obvious transverse character, should show no holonomy. This is not observed experimentally \cite{Tomita86}. Although it might be possible to invoke the influence of optical fiber geometry on rotation of the polarization plane, considerations based on the orientation of the field with respect to time allow an alternate interpretation.  

The classical sum of helicity states described in the preceding paragraph is \cite{Baylis99} chap. 7
\begin{equation}\label{classical_phase}
e^{\imag\, p_\mu x^\mu}|+\rangle + e^{-\imag\, p_\mu x^\mu}|-\rangle
\end{equation}
where for avoidance of doubt we can multiply both states by the same modulation envelope, such as a simple Gaussian functional $(2\pi\sigma^2/p_0^2)^{-1/4}\exp(- (p_\mu x^\mu)^2/(4 \sigma^2))$ \cite{Baylis99}, and confirm that the resulting structure describes a plane wavelet propagating forward in time. (\ref{classical_phase}) has no obvious pathologies from a classical perspective. However it does not allow a Hamiltonian equation that is consistent with all properties of the field, even if the sum is taken to be a direct tensor addition, because the derivative with respect to time yields both positive and negative energies and the resulting currents are then pseudovectors, not vectors. Consistent with this, if (\ref{classical_phase}) is multiplied by a Gaussian modulation function the Fourier transform of the resulting state is a sum of two momentum amplitude components, one centered on $p$ and the other on $-p$. If we require a common orientation for all states in both time and time derivative, we have a direct sum of the form (\ref{direct_sum}), and the Hamiltonian has the form
\begin{equation}
\mathcal{H}=\left(\begin{matrix} \mathcal{H}_+ & \mathbf{0}_4 \cr \mathbf{0}_4 & \mathcal{H}_-  \cr  \end{matrix}\right)= \left(\begin{matrix} \chi_j p_j & \mathbf{0}_4 \cr \mathbf{0}_4 & -\chi_j p_j  \cr  \end{matrix}\right)
\end{equation}
which has quadruply degenerate eigenvalues of $\pm p_0$. The time evolution of states is given by
\begin{equation}
\left(\begin{matrix} e^{-x_0 \mathcal{H}_+} & \mathbf{0}_4 \cr \mathbf{0}_4 & e^{-x_0 \mathcal{H}_-}  \cr  \end{matrix}\right)\,\left(\begin{matrix} \psi_+\cr -\psi_-\cr \end{matrix}\right)\, e^{\imag\, \mathbf{p\cdot x}} = \left(\begin{matrix} \psi_+\cr -\psi_-\cr \end{matrix}\right)\, e^{\imag\, p_\mu x^\mu}
\end{equation}
With this, multiplication of $\psi_+$ and $\psi_-$ by $e^{\imag\, \gamma}$ and $e^{-\imag\, \gamma}$, respectively, is equivalent to phase translation in opposite time directions, leading to rotation of the plane of polarization. Representing the polarization state by the parameters of the Poincar\'{e} spinor (helicity basis), we have for the positive energy part of the field
\begin{equation}
\left(\begin{matrix} \psi_+ \, a_+\cr -\psi_- \, a_-\cr \end{matrix}\right)\, e^{\imag\, p_\mu x^\mu}
\end{equation}
where $a_+$ and $a_-$ are the annihilation operators for the positive and negative helicity states, respectively. In the standard depiction, the `north' and `south' poles of the Poincar\'{e} sphere are pure $+$ and $-$ helicity states represented by amplitude weights of $\cos \theta/2$ and $e^{-\imag \,\phi}\sin \theta/2$ respectively, and the equatorial antipodes (front and back) are pure `x' and `y' states. The plane $\theta = \pi/2$ is the plane of linear polarization, and $\theta = \pi/2, \phi=0$ corresponds to linear x axis polarization while $\theta = \pi/2, \phi=\pi$ corresponds to linear y axis polarization. From this we can see that the effect of multiplication of an equally weighted mixture of positive and negative states ($\theta = \pi/2$) by $e^{\imag\, \gamma}$ and $e^{-\imag\, \gamma}$, respectively, is equivalent to a rotation of the plane of linear polarization by $\gamma$; a corollary is that the phase factors of (\ref{rotation_phase}) do not have a simple geometric interpretation. With respect to the angular momentum dependence found by Berry, we must also conclude that the equally weighted direct sum of time-oriented $(1,0)$ and $(0,1)$ representations does not transform as a state of zero angular momentum under rotations.

In the standard spherical coordinates $p_1 = p_0 \cos \phi \sin\theta, p_2 = p_0 \sin \phi \sin\theta, p_3 = p_0\cos \theta$, (\ref{rotation_phase}) gives 
\begin{equation}
e^{\pm \imag \arctan (\cos\theta \tan \phi)} \psi_\pm
\end{equation}
and by the Gauss-Bonnet theorem, for a large class of simple loops that enclose an area that can be represented as the union of a countable collection of triangles, the exponent is equal to the area of the spherical surface enclosed divided by the square of the radius of curvature, $|p|$. For the calculation above this is easily confirmed in the limit $\theta \rightarrow 0$. The general result can be reached in a direct and elegant way from the properties of the holonomy group and the unitary representation of the massless spin one little group \cite{Lindner03}. 

Reversal of the direction by which the path is traversed leads to reversal of the sign of the exponent. In addition it can easily be seen that for a spin $n$ field, constructed as the outer product of $n$ spin 1 fields, the phase acquired upon traversal of the circuit would be $n$ times the value obtained upon traversal by a spin one field (apply the sequence of transformations to each element of the outer product). The related $SU(2)$ phase value $\pm\arctan (\cos\theta \tan \frac{\phi}{2})$ \cite{Wagh98,Sjoqvist01} has been identified with the Pancharatnam phase \cite{Pancharatnam56}, obtained by rotating the polarization state about a closed path on the Poincar\'{e} sphere.

Rotation of the vector solutions of (\ref{eigenvectors1}) to an arbitrary direction of motion by recapitulation of the steps leading to (\ref{z2} - \ref{z1}) is simple and yields the same result in all three cases,
\begin{equation} \label{z_T}
z_T(p) := \frac{1}{{\sqrt{2}}\, {p_0}}\left( \imag\, {p_0}, {p_1}, {p_2}, {p_3}\right) 
\end{equation} 
the longitudinal field, with time-like polarization. There is no phase freedom in this case because the solutions describe polar, as opposed to tangent, vectors; there is correspondingly no quantization known to be associated with the longitudinal field. 

The prescription for boost transformation of $z_T$ differs from that for $z_S$, and as was the case for the $z_S$, the transformation laws for $z_T$ and $\overline{z}_T$ are different:\begin{equation}\label{z_T_boost}
z_T(p^\prime) = e^{\,\kappa \, K_i} z_T(p) = z_T(p)  e^{-\kappa \, K_i}
\quad\quad \overline{z}_T(p') = e^{-\kappa \, K_i} \overline{z}_T(p) = \overline{z}_T(p) e^{\, \kappa \, K_i} \end{equation}
The rotation operator is the same for $z_S$ and $z_T$ (and $\overline{z}_S$ and $\overline{z}_T$)
\begin{equation}\label{z_T_rotation}
z_T(p^\prime) = e^{\,\imag \kappa \, S_j} z_T(p)  = z_T(p) e^{-\imag \kappa \, S_j} 
\end{equation}

\section{Phase Freedom and Quantization}
Phase freedom imposes additional constraints on field structure. Because the wave-like behavior of elementary particles is experimentally well-validated there must be some principle that prevents the phase of the field from changing arbitrarily from one space-time coordinate to another. If this were not present all phases could equally coexist at all space-time coordinates -- or equivalently be unpredictable -- and interference phenomena would not be observable. The simplest reconciliation of phase freedom with wave-like behavior is to posit that the phase of a photon is determinate, but changes upon rotation from one coordinate system to another in a path-dependent manner. That is, that phase is embedded topologically and depends on the rotation history of both the particle and the observer. 

Now imagine that the photon field were not quantized, specifically that the angular momentum could be varied continuously by some unspecified classical mechanism. The angular momentum in this hypothetical field would presumably be represented by an intensive variable, i.e., would be localizable. Since the phase change associated with variation in path orientation depends linearly on angular momentum, frame rotation of a wave train would result in a phase change proportional to the total angular momentum accumulated at the point of measurement, in turn dependent on the length of the electromagnetic impulse measured. We could formalize this by saying the field would have an additional multiplicative factor $e^{\imag\, x_0 \theta}$ where $\theta$ is proportional to the solid angle of the circuit traversed (in the case of a closed loop). However this would be equivalent to a change in energy of the system by $-\hbar\, \theta$. If this were not sufficiently disquieting, by the analysis above, the plane of linear polarization would rotate with time.  Although it has been argued \cite{Haldane87,Segert87} that the rotation of the plane of polarization of linearly polarized light injected into a helically wound optical fiber \cite{Chiao86,Tomita86} represents a classical effect, the necessity for a discrete angular momentum, and hence field quantization, appears unavoidable.  

\section{Current and Gradient Operators}
For massive and massless spin 1/2 fields, there exist well behaved four-vector current operators whose expectations, typically written $\overline{\psi}\gamma_\mu\psi$ and $\psi^\dagger\sigma_\mu\psi$, respectively, yield the predicted particle flux. If these did not exist there would be difficulties accounting for conservation of charge. Because photons do not bear charge, there is no obvious necessity for a conserved current, but there are related requirements to preserve conservation of energy-momentum. The standard definition of the quantized quantity corresponding to the Poynting energy-momentum flux vector is an expectation over a field and the time derivative of its adjoint: 
\begin{equation}\label{standard_current}
p^{\,\mu}=-\int d^3 x\dot{A}^{(-)}(x,t) \partial/\partial x^{\,\mu} A^{(+)}(x,t)
\end{equation}
This has been formally justified in the context of the canonical quantization of boson fields, which are thought to have canonical adjoints that are time derivatives of the ordinary adjoints; the canonical adjoint of spinor fields is the ordinary adjoint. Depending on one's perspective, either quantization, boson or fermion, could appear unnatural. For example, in the fermion case the canonical momentum is the adjoint field. 

From field equations of the form $\partial_\mu \,\chi^{\,\mu} \psi_+ = 0$, it would be reasonable to attempt to identify the $\chi^j$ with current operators, yielding operator expectations of the form $ \int d^3 x \,[ {\psi_+}^\dagger \chi^{\,\mu} {\psi_+}] = j^{\,\mu} $, while allowing translation operator expectations to be formed in the usual way: $\int d^3 x \,[{\psi_+}^\dagger \partial_\mu \,\psi_+] = p^{\,\mu}$. Although such a representation can be realized, the current operators transform not as vectors but as tensors, because the fields themselves are tensors. 

Our first task is to introduce an adjoint field as a formal conjugate, written in terms of the electric and magnetic amplitudes as
\begin{equation} 
{\psi_+}^\dagger = \frac{e^{-\imag \,\theta}}{\sqrt{2p_0}}( 0,-\imag \,{b_1}^\dagger + {e_1}^\dagger,-\imag \,{b_2}^\dagger + {e_2}^\dagger,-\imag \,{b_3}^\dagger + {e_3}^\dagger ) 
\end{equation}
In the presence of interactions the adjoint field is an independent quantity and by convention labels the exiting particle. The momentum function conjugates corresponding to this field satisfy the adjoint equation $\overline{z}(p) p^{\,\mu} \,\chi_\mu = 0$ as well as the contravariant (negative helicity) equation $p^{\,\mu} \chi^{\,\mu} \,\overline{z}(p) = 0$. The latter correspond to the Maxwell equations 
\begin{equation} 
\partial^{\,\mu} \chi^{\,\mu} {\psi_+}^\dagger = (\mathbf{\nabla \cdot b^\dagger} +\imag \mathbf{\nabla \cdot e^\dagger}, -\partial_0 \mathbf{e^\dagger} + \mathbf{\nabla \times b^\dagger} + \imag (\partial_0 \mathbf{b^\dagger} + \mathbf{\nabla \times e^\dagger} ))\frac{e^{-\imag \,\theta}}{\sqrt{2p_0}} = (\imag \rho^\dagger, \mathbf{j}^\dagger)\frac{e^{-\imag \,\theta}}{\sqrt{2p_0}}
\end{equation}
where we have superscripted the exiting current to signify its possibly different structure from that associated with the entering field.
 
The Poynting currents for the field can be calculated as elements of the symmetrical tensor formed from the field and adjoint, expressed in components as
\begin{equation} \label{psi_stress_energy}
{\psi_\pm}^{\dagger \,\lambda} {\chi^{\,\alpha}}_{\lambda \mu} \chi^{\beta \mu \nu} \psi_{\pm\,\nu}
\end{equation}
The currents formed from the amplitudes can be written
\begin{equation} \label{momentum_current}
{\psi_\pm}^\dagger \chi^{\,0} \chi^{\,i} {\psi_\pm} = {\psi_\pm}^\dagger \chi^{\,i} {\psi_\pm} =\pm{\epsilon}_{ij\,k}\,
   \left({e_j}^{\dagger}\,{b_k} - {b_j}^{\dagger}\,{e_k} \mp
     \imag \,\left( {b_j}^{\dagger}\,{b_k} + {e_j}^{\dagger}\,{e_k} \right)  \right)/(2 p_0)
\end{equation}
and 
\begin{equation} \label{energy_current}
 {\psi_\pm}^\dagger \chi^{\,0} \chi^{\,0} {\psi_\pm} =  ({b_j}^\dagger b_j + {e_j}^\dagger e_j\pm 
\imag \,({e_j}^\dagger  b_j-  {b_j}^\dagger e_j ) )/(2 p_0) 
\end{equation}

The spatial elements of the stress-energy tensor are
\begin{eqnarray}\label{stress_energy}
 {\psi_\pm}^\dagger \chi^{\,l} \chi^{\,m} {\psi_\pm} = \delta_{lm}({b_j}^\dagger b_j + {e_j}^\dagger e_j\pm 
\imag \,({e_j}^\dagger  b_j-  {b_j}^\dagger e_j ) )\nonumber\\
 - ({b_l}^\dagger b_m +{b_m}^\dagger b_l + {e_l}^\dagger e_m+ {e_m}^\dagger e_l\pm 
\imag ({e_l}^\dagger  b_m-  {b_m}^\dagger e_l +{e_m}^\dagger  b_l-  {b_l}^\dagger e_m) )
\end{eqnarray}

In the absence of interaction the field component operators ${b_i}^\dagger ,  e_j$ etc., must form Hermitian products (i.e., ${b_i}^\dagger e_j = {e_j}^\dagger b_i$) to insure that the momenta, etc., are real. With this the imaginary terms in (\ref{momentum_current}-\ref{stress_energy}) vanish and 
\begin{equation}
{\psi_\pm}^\dagger \chi^{\,i} {\psi_\pm} = \pm{\epsilon}_{ij\,k}\,
   \left({e_j}^{\dagger}\,{b_k} - {e_k}^{\dagger}\,{b_j} \right)/(2 p_0)
\end{equation} 
and 
\begin{equation} {\psi_\pm}^\dagger \chi^{\,0} {\psi_\pm} = ({b_j}^\dagger b_j + {e_j}^\dagger e_j)/(2 p_0) 
\end{equation}
which reduce to the classical expressions. The Hermiticity constraint does not apply in the presence of interactions, as the incoming and outgoing fields have different energy-momentum structures. 

In the absence of interaction the current bilinears generated from the momentum functions of the fields and adjoints have the properties of a generalized velocity, i.e. 
\begin{equation}
{\overline{z}_T(p)} \,\chi^{\,\mu} {z_T(p)} =\overline{z}_S(p)\, \chi^{\,\mu} {z_S(p)} = p^{\,\mu}/p^0
\end{equation}
although it should be kept in mind that in the case of the $z_S$, the currents are components of a tensor, not a vector. For example the full relation for the stress-energy tensor is
\begin{equation}
\overline{z}_S(p)\,{\chi^{\,\alpha}}_{\lambda \mu} \chi^{\beta \mu \nu} {z_S(p)} = p^{\,\alpha}p^{\,\beta}/p_0^2
\end{equation}
and the action of the boost operator on the currents is a tensor reflection,
\begin{equation} \label{field_boost}
j^{\,\mu\,\prime} = \psi^\dagger e^{\,\kappa \,S} \chi^{\,\mu} e^{\,\kappa \,S} \psi 
\end{equation}
whereas the action of the rotation operator is a tensor rotation
\begin{equation}  \label{field_rotation}
j^{\,\mu\,\prime} = \psi^\dagger e^{-\imag\,\kappa \,S} \chi^{\,\mu} e^{\,\imag\,\kappa \,S} \psi 
\end{equation}
The boost operation is a reflection because the current operator corresponding to the time dimension is a scalar; if it were a vector element (the usual case) the boost operation would be a rotation.

The field expectation of translation operators is, by most definitions, computed over the scalar part of the field. In the case of a scalar particle, there is no current operator known apart from the translation operator, and the definition 
\begin{equation}\label{translation_operator}
p^{\,\mu}=\frac{-\imag}{2}\int d^3 x \,:[{\psi_+}^\dagger \partial_\mu \,\psi_+ -\psi_+ \partial_\mu \,{\psi_+}^\dagger]:  
\end{equation}
where the colons denote normal ordering of creation and annihilation operators, can be identified with the operator expectation. This definition extends to both spinor and bivector fields, and to higher order fields for which the adjoint can be identified as the field quantity that yields a probability density upon contraction with the field. 

\section{Orthogonality Relations}
The $z_T$ and $z_S$ are orthonormal with respect to inner product (with the application of the energy-momentum constraint $p_\mu p^{\,\mu}=0$):
\begin{eqnarray}
\overline{z}_T \,z_T   = \overline{z}_S \,z_S  = 1\nonumber\\
\overline{z}_T \,z_S = \overline{z}_S \,z_T = 0
\end{eqnarray}
and the electric and magnetic amplitudes in the transverse representation have the additional normalization and orthogonality properties (with the energy-momentum constraint)
\begin{eqnarray} \label{electromagnetic_amplitude_properties}
ze(p)\,zb(p) = & 0 \nonumber\\
z_T(p)\,ze(p) = z_T(p)\,zb(p) = & 0 \nonumber\\
ze(p)\,ze(p) = zb(p)\,zb(p) = & 1/2
\end{eqnarray}
consistent with the idea that an orthogonal basis can be formed from vectors parallel to the momentum, the electrical field and the magnetic field, and that the electric and magnetic components of a source-free simple field contribute equally to the energy density of the field.

\section{Projection operators}
Projection of positive and negative energy states requires an idempotent operator, usually formed as a sum over (orthogonal) spin states of the normalized outer product of basis vectors and their adjoints, $P = \sum_\sigma \,\psi \otimes \psi^\dagger$. In the representation developed here the sum is not over spin states but rather longitudinal and transverse components.
\begin{eqnarray}
z_S \otimes \overline{z}_S = & {\delta_{i\,j}}\,/{2} -{p_i\,p_j}\,/{2 {p_0}^2} -\imag \epsilon_{i\,j\,k}{p_k}\,/{2 p_0} & 1\leq i,j \leq 3 \nonumber \\
 = & 0 & \textrm{otherwise} 
\end{eqnarray}
The spatial components of $z_T \otimes \overline{z}_T$ are ${p_i\,p_j}\,/{2 {p_0}^2}$, leading to
\begin{equation} \label{pos_proj_operator}
P_+ = z_T \otimes \overline{z}_T + z_S \otimes \overline{z}_S = \frac{p^{\,\mu}\chi^{\,\mu}}{2 p_0} 
\end{equation}
\begin{equation} \label{neg_proj_operator}
P_- = \overline{z}_T \otimes z_T +  \overline{z}_S \otimes z_S = \frac{- p^{\,\mu}\chi_\mu}{2 p_0} 
\end{equation}

These have the required properties $P_\pm\,P_\pm = P_\pm$ and $P_\pm\,P_\mp = 0$. Also $P_+ \,z = z; P_- \,\overline{z} = \overline{z}; P_+ \,\overline{z} = P_- \,z = 0$. Because the negative energy solutions are indistinguishable from lefthanded solutions in the case of the transverse field, these operators also project out spin states. The structure of the projection operators illustrates the principle that both transverse and longitudinal contributions must be included to form a complete description of the field. That both contributions are required is also implicit in the description of the Maxwell equations. 

\section{Connection with other formalisms}
In some treatments the relation $-\partial A/\partial t = E$ (in radiation gauge) is taken to justify the representation of the photon in terms of a quantized electric field
\begin{equation}
E^{\,\mu}(x)= (2\pi)^{-3/2} \sum_\sigma \int^\infty_{-\infty} d^3 p\,\sqrt{\frac{p_0}{2}}[e^{\mu}(\mathbf{p},\sigma) a(\mathbf{p},\sigma)e^{\imag p x} + e^{\mu *}(\mathbf{p},\sigma) a^\dagger(\mathbf{p},\sigma)e^{-\imag p x}]
\end{equation}
in which the polarization vectors $e^{\mu}(\mathbf{p},\sigma)$ are the same as those used to describe the vector potential. The magnetic amplitudes are given by 
\begin{equation}
H^{\,\mu}(x) = (2\pi)^{-3/2} \sum_\sigma \int^\infty_{-\infty} d^3 p\,\sqrt{\frac{p_0}{2}}[\frac{(\mathbf{p}\times e(\mathbf{p},\sigma))^{\,\mu}}{p_0} a(\mathbf{p},\sigma)e^{\imag p x} + \frac{(\mathbf{p}\times e(\mathbf{p},\sigma))^{\mu *}}{p_0} a^\dagger(\mathbf{p},\sigma)e^{-\imag p x}]
\end{equation}
This depiction inherits some of the difficulties of the vector potential representation. The electromagnetic field transforms not as a vector (\ref{z_T_boost}) but as a tensor (\ref{z_S_boost}); the polarization bivectors $e(\mathbf{p},\sigma)$ describe the complex sum of electric and magnetic fields, not the individual fields in isolation; and the field conjugates transform by different laws and obey distinct field equations. 

The approach taken in this work appears to be most closely related to that of Bialynicki-Birula and Bialynicka-Birula \cite{Bialynicki-Birula87} who have emphasized the role of the Poincar\'{e} group in determining the structure of the fields. A gauge dependent covariant derivative operator does not feature in this analysis, however -- the comparable action is fulfilled by operator inner products of the form (\ref{covariant_equation}-\ref{contravariant_equation}) in the momentum representation.

\section{Modulation Envelope Terms}
A vector potential with current given by (\ref{standard_current}), and a bivector field with current based on (\ref{momentum_current}) or (\ref{translation_operator}) can be distinguished by the presence or absence of terms dependent on the modulation envelope. Introducing the generic real envelope functional $f(p_\mu x^{\,\mu})$, subject to the constraint $\int f^2 d^3 x =1$ we note that, as in the classical case \cite{Baylis99}, $\psi = f(p_\mu x^{\,\mu}) e^{\imag p_\mu x^{\,\mu}}$ is a solution of the wave equation $\partial^{\,\mu} \partial_\mu \psi = 0$. Taking a pure polarization state, for example pure positive helicity light propagating in the 1-direction in a discrete single mode with wavenumber $k$, the vector potential wavelet
\begin{equation}
A(x^{\,\mu}) = \frac{f(k_\mu x^{\,\mu})}{\sqrt{2 \omega_k}} (\,e(k,\sigma)  e^{\imag k_\mu x^{\,\mu}} a_\mathbf{k} + e(k,\sigma)^* e^{-\imag k_\mu x^{\,\mu}} {a_\mathbf{k}}^\dagger := A^+(x^{\,\mu}) + A^-(x^{\,\mu})
\end{equation}
gives, by (\ref{standard_current}), a field momentum
\begin{equation}\label{vector_current}
{k_A}^{\,\mu} = -\int d^3 x [\partial_0 A^-(x^{\,\nu}) \partial^{\,\mu} A^+(x^{\,\nu})] = {a_\mathbf{k}}^\dagger {a_\mathbf{k}}\int d^3 x[k^{\,\mu} (f^2 + {f^{\,\prime}}^2)]={a_\mathbf{k}}^\dagger {a_\mathbf{k}}\, k^{\,\mu} (1+\int d^3 x[{f^{\,\prime}}^2])
\end{equation}
whereas the bivector field wavelet
\begin{equation}
B(x^{\,\mu}) = f(k_\mu x^{\,\mu}) (e(k,\sigma) e^{\imag k_\mu x^{\,\mu}} a_\mathbf{k} \oplus e(k,\sigma)^* e^{-\imag k_\mu x^{\,\mu}} {a_\mathbf{k}}^\dagger := B^+(x^{\,\mu}) \oplus B^-(x^{\,\mu})
\end{equation}
gives, by (\ref{translation_operator}), a field momentum
\begin{equation}\label{bivector_current}
{k_B}^{\,\mu} = \frac{-\imag}{2} \int d^3 x :[B^-(x^{\,\nu}) \partial^{\,\mu} B^+(x^{\,\nu})-B^+(x^{\,\nu}) \partial^{\,\mu} B^-(x^{\,\nu})]:\,\, = {a_\mathbf{k}}^\dagger {a_\mathbf{k}}\, k^{\,\mu}
\end{equation}
The difference between (\ref{bivector_current}) and (\ref{vector_current}) is the spatial integral of a squared functional derivative. To see what this contributes we evaluate it for the case of the normalized plane wavelet envelope
\begin{equation}\label{gaussian_wavelet}
f(k_\mu x^{\,\mu}) = (2 \pi \sigma^2/k_0^2)^{-1/4} e^{-\frac{(k_\mu x^{\,\mu})^2}{4\sigma^2}}
\end{equation}
specialized to a pure 1-direction motion (integration over the 2- and 3- directions is taken to give unity). In this case the Fourier transform in the two spatial variables $x_0,\,x_1$ of the positive frequency part of the field gives 
\begin{equation}
\int f(k_\mu x^{\,\mu})e^{\imag k_\mu x^{\,\mu}}\,e^{-\imag p_0 x_0 +p_1 x_1} dx_0\, dx_1  = (8 \pi \sigma^2/k_0^2)^{1/4}e^{\frac{-\sigma^2\,(p_1-k_1)^2}{k_1^2}}\delta(p_0-p_1)
\end{equation}
and the squared derivative term in (\ref{vector_current}) gives  
\begin{equation}\label{gaussian}
\int  dx_1 [{f^{\,\prime}}^2] = \int  dx_1 [\frac{k_0\,(k_0x_0 -k_1x_1)^2}{2^{5/2}\sigma^5\pi^{1/2}}e^{-\frac{(k_0 x_0 -k_1 x_1)^2}{2\sigma^2}}] = \frac{1}{4\sigma^2}
\end{equation}
so that the current (\ref{vector_current}) is
\begin{equation}\label{vector_current_1}
{k_A}^{\,\mu} = {a_\mathbf{k}}^\dagger {a_\mathbf{k}}k^{\,\mu}(1 + \frac{1}{4\sigma^2 })
\end{equation}
An energy-momentum contribution that increases with decreasing width of the modulation envelope should be demonstrable experimentally. 

\section{Currents from Field Superposition}
Another setting that allows discrimination of vector and bivector fields is the prediction of currents arising from field superposition. Consider for instance a field that is constituted by the division and redirection of two components. Let the two components be directed toward each other symmetrically after some experimental manipulation to separate them, and for convenience, let them be in the same polarization state. If the envelope functional of one component is $f(k_\mu x^{\,\mu})$ and of the other is $g(-k^\mu x^{\,\mu})$ (i.e., they are directed with equal and opposite momenta, evolving in the same time direction), and if we set the time of intersection of the fields to be the origin, the positive frequency part of the bivector superimposed field will be
\begin{equation}\label{superimposed_bivector_field}
 B^+(x^{\,\mu})=  f(k_\mu x^{\,\mu}) e(k,\sigma) e^{\imag k_\mu x^{\,\mu}} a_\mathbf{k}+ g(-k^\mu x^{\,\mu}) e(k,\sigma) e^{-\imag k^\mu x^{\,\mu}} a_\mathbf{k}
\end{equation}
in which the fields share the same creation and annihilation operators because they are derived from a single source. Using (\ref{superimposed_bivector_field}) and (\ref{translation_operator}) we find for the spatial components of the momenta
\begin{equation}\label{superimposed_bivector_current}
<{k_B}^{\,j}> = {a_\mathbf{k}}^\dagger {a_\mathbf{k}}\, k^{\,j}\int d^3 x [f^2 - g^2+2 Sin(2\, \mathbf{p\cdot x})(f^{\,\prime}g+g^{\,\prime}f)]
\end{equation}
where the subtraction of squared functionals reflects their opposite contributions to total momentum; for the energy we find
\begin{eqnarray}
<{k_B}^{\,0}> = {a_\mathbf{k}}^\dagger {a_\mathbf{k}}\, k^{\,0}\int d^3 x [f^2 + g^2+2fg\, Cos (2\,\mathbf{p\cdot x})
+2 Sin(2 \,\mathbf{p\cdot x})(f^{\,\prime}g-g^{\,\prime}f)]
\end{eqnarray}
so the energy is given by the sum of the energies of the components, as expected. For the case of a pure 1-direction motion, with a Gaussian envelope of the form (\ref{gaussian_wavelet}) for $f$, and with  $g$ given by the same formula with opposite sign of $p_1$, the interference term in (\ref{superimposed_bivector_current}) is given by 
\begin{equation}
\frac{1}{{\sqrt{2\,\pi \,{\sigma }^2}}}\,
   \frac{e^{-\frac{{{p_0}}^2\,{{x_0}}^2 + {{p_1}}^2\,{{x_1}}^2}{2\,{\sigma }^2}}}{{\sigma }^2}\,{p_0^2}\,{x_0}\,
   \sin (2\,{p_1}\,{x_1})
\end{equation}
Analysis of the currents by the vector potential treatment gives 
\begin{eqnarray}\label{superimposed_vector_current}
<{k_A}^{\,\mu}> = {a_\mathbf{k}}^\dagger {a_\mathbf{k}}\, k^{\,\mu}\int d^3 x [f^2 +{f^{\,\prime}}^2 - g^2 -{g^{\,\prime}}^2 +2\,\imag \,Sin(2 \mathbf{p\cdot x})(f^{\,\prime}g+f\,g^{\,\prime})
+ 2\,\imag\, Cos (2 \mathbf{p\cdot x})(f g^{\,\prime}-f^{\,\prime}g)]\qquad
\end{eqnarray}
hence in addition to the squared functional derivatives encountered in (\ref{vector_current}) there are imaginary interference terms. 

\bigskip
I thank Jeremy Gunawardena, Tom McCauley, and Rebecca Ward for suggestions and encouragement of this work. 


\end{document}